\documentclass[twocolumn, preprintnumbers,amsmath,amssymb,prl]{revtex4}

\usepackage{epsfig}
\usepackage{dcolumn}
\usepackage{bm}
\usepackage{latexsym}

\begin{document}

\hyphenation{thres-hold he-te-ro-struc-tures}

\title{Vertical cavity surface emitting laser action of an all monolithic ZnO-based microcavity}

\author{S. Kalusniak}
\author{S. Sadofev}
\author{S. Halm}
\email{simon.halm@physik.hu-berlin.de}
\author{F. Henneberger}

\affiliation{Institut f\"{u}r Physik, Humboldt-Universit\"{a}t zu Berlin, Newtonstr. 15, 12489 Berlin, Germany}

\date{\today}

\begin{abstract}

{We report on room temperature laser action of an all monolithic
ZnO-based vertical cavity surface emitting laser (VCSEL) under
optical pumping. The VCSEL structure consists of a 2$\lambda$
microcavity containing 8 ZnO/Zn$_{0.92}$Mg$_{0.08}$O quantum wells
embedded in epitaxially grown
Zn$_{0.92}$Mg$_{0.08}$O/Zn$_{0.65}$Mg$_{0.35}$O distributed Bragg
reflectors (DBRs). As a prerequisite, design and growth of high
reflectivity DBRs based on ZnO and (Zn,Mg)O for optical devices
operating in the ultraviolet and blue-green spectral range are
discussed.}

\end{abstract}


\maketitle

Semiconductor lasers operating in the ultraviolet (UV) and blue
spectral range have become of great technological importance, e.g.,
for data storage applications. Compared to the commonly employed
edge-emitting devices, vertical cavity surface emitting lasers (VCSELs) offer several advantages such as
single mode emission, a circular beam profile, and the possibility
for integration into two dimensional arrays. VCSELs based on GaN and
its alloys reached a promising stage of development \cite{Shimada10}
and laser devices working in the strong exciton-photon coupling
regime, so called polariton lasers, have been demonstrated at room
temperature \cite{Christopoulos07}. In principle, with ZnO-based
alloys, a similar spectral range can be covered as with the (Al,In,Ga)N
heterosystem. Room temperature lasing of (Zn,Cd)O/ZnO quantum well
(QW) structures was achieved from UV to green wavelengths
\cite{Kalusniak09}. In the strong coupling regime, ZnO was predicted
to be even superior due to its high oscillator strength and exciton
binding energy \cite{Zamfirescu02,Faure08}. However, necessary
prerequisites to both VCSELs and polariton lasers are microcavities
(MCs) with a high quality factor and good optical quality ZnO active
layers. So far only hybrid MCs
\cite{Shimada08,Chen09,Faure09,Sturm09} consisting of, e.g., an
epitaxially grown GaN-based bottom distributed Bragg
reflector (DBR), a ZnO active layer, and a top
dielectric DBR have been employed.

In this letter, we report on the monolithic growth of ZnO and
(Zn,Mg)O based DBRs and MCs by radical-source molecular beam epitaxy (MBE). With a properly chosen DBR ternary
alloy composition, it is possible to obtain high reflectivities with
a moderate number of mirror pairs in the range from $375$~nm to
$500$~nm. Finally, we demonstrate an UV emitting all-ZnO VCSEL.

The samples are grown in a commercial DCA-450 MBE apparatus equipped with standard metal
sources and a radical plasma cell for oxygen. A-plane ($11\bar{2}0$) sapphire wafers are used as substrates. The (Zn,Mg)O DBRs and MCs
containing ZnO/(Zn,Mg)O QWs are grown without interruption at $T_\textnormal{g} = 340$~$^\circ$C with an oxygen limited growth
rate of 330~nm/h, which was previously determined from oscillations in the specular beam intensity of \textit{in situ} reflected high-energy electron diffraction (RHEED) \cite{Kalusniak09}. Note that the low $T_\textnormal{g}$  allows for the growth of both $\mu$m-thick, high Mg-content (Zn,Mg)O layers with a surface roughness of only a few monoatomic layers as well as high optical quality ZnO QWs \cite{Sadofev05}.
Individual layer thicknesses are defined via growth times. Two effusion cells for Mg are used in order to modulate the alloy composition of ternary Zn$_{1-x}$Mg$_x$O layers. The overall structure quality is continuously monitored by detecting the RHEED pattern of the growing surface.

\begin{figure}
 \epsfxsize=8.5cm
 \centerline{\epsffile{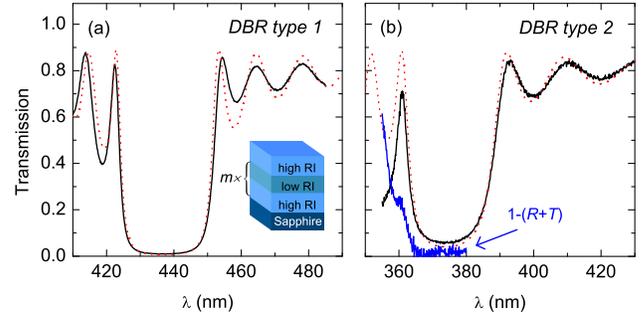}}
 \caption{\label{fig:1} (Color online) Optical properties of ZnO-based DBRs type 1 and 2. Transmission spectra (solid lines) and calculated transmission (dotted lines) of a ZnO/Zn$_{0.65}$Mg$_{0.35}$O DBR (a) and a Zn$_{0.92}$Mg$_{0.08}$O/Zn$_{0.65}$Mg$_{0.35}$O DBR (b). Inset in (a): Schematic layout (not true to scale). In (b), absorption $A=1-(R+T)$ is added as reference (see arrow).}
\end{figure}

Two types of ZnO-based DBRs have been realized:
ZnO/Zn$_{1-x}$Mg$_{x}$O (type 1) and
Zn$_{1-y}$Mg$_{y}$O/Zn$_{1-x}$Mg$_{x}$O (type 2) for blue emitting
(Zn,Cd)O and UV emitting ZnO optical devices, respectively.
Figure~\ref{fig:1}(a) shows the transmission spectrum of a type-1
DBR with $m=27$ pairs of $\lambda/4$-layers of
ZnO/Zn$_{0.65}$Mg$_{0.35}$O and an additional $\lambda/4$-thick ZnO
cap layer (schematic layout: see inset). A Mg content of $x=0.35$ slightly below the solubility
limit was selected for the low refractive index (RI) layer
\cite{Sadofev05} in order to reach a large RI contrast ($\Delta
n/n\approx 10\%$) while still allowing for stable two-dimensional
growth. For the specific layer thicknesses used, the stop band is centered
at $\lambda_\textnormal{SC}=436$~nm and exhibits a spectral width of
about 180~meV. The number of mirror pairs is sufficient to reduce
the transmission through the stack at $\lambda_\textnormal{SC}$ to $T=0.01$.
In type-2 DBRs aimed for ZnO devices, the high RI material has to contain a
few percent of Mg to avoid absorption in the mirrors. We found that
$y=0.08$ is a good choice which allows to keep $\Delta n/n$ as large
as possible. The transmission of a $m=15$ pairs
Zn$_{0.92}$Mg$_{0.08}$O/Zn$_{0.65}$Mg$_{0.35}$O-DBR with an
additional $\lambda/4$-Zn$_{0.92}$Mg$_{0.08}$O layer is depicted in
Fig.~\ref{fig:1}(b). At the stop band center
$\lambda_\textnormal{SC}=375$~nm, the transmission is reduced to
$T=0.055$. The transmission values can be directly converted into the DBR reflection $R=1-T$ if absorption is negligible. We have ensured this by recording complementary reflectivity spectra. For demonstration, absorption $A=1-(R+T)$ is shown in Fig.~\ref{fig:1}(b) (see arrow) for the type-2 DBR, where $\lambda_\textnormal{SC}$ is closest to the high RI layer absorption onset. In fact, the presence of spectrally sharp absorption edges, also for ternary (Zn,Mg)O layers, is an essential feature for the DBR functionality we report on here. We find thus $R=0.99$ and $R=0.945$ for the type-1 and type-2 DBRs in Fig.~\ref{fig:1}, respectively.

To model the experimental data by transfer matrix calculations \cite{Born99}, we describe the RI dispersion by a first-order
Sellmeier equation

\begin{equation}
n(\lambda)=\sqrt{A +\frac{B\lambda^{2}}{\lambda^{2}-C^{2}}},
\label{eqn:1}
\end{equation}

\noindent and neglect absorption [see dotted lines in Fig.~\ref{fig:1}(a) and (b)]. For
ZnO, where $n(\lambda)$ is well-established, the parameters $A$, $B$
and $C$ shown in Table~\ref{tab:1} result from a least-square fit of
Eq.~(\ref{eqn:1}) to the data of Ref. \cite{Mollwo54}. For (Zn,Mg)O,
RI literature data still differ strongly and cover only part of the
wavelength range of interest here \cite{Teng00,Schmidt03}. Our
values for $x=0.08$ and $x=0.35$ in Table~\ref{tab:1} correspond to
a ``best guess'' for which the transfer matrix model reproduces the transmission spectra of five experimentally realized
DBRs very well (type~1: $\lambda_\textnormal{SC}=422$~nm and $436$~nm;
type 2: $\lambda_\textnormal{SC}=375$~nm, $382$~nm, and
$394$~nm). The resulting dispersion curves are in reasonable
agreement with Ref.~\cite{Teng00}. The good match between model and experiment in the stop band region further confirms that absorption is not important here.

\begin{table}[b]
 \caption{Sellmeier values $A$, $B$ and $C$ for DBR layers with Mg content $x$.}
 \begin{ruledtabular}
  \begin{tabular}{cccc}
             $x$ & $A$ & $B$ & $C$ (nm) \\ \hline
             0.00 & 3.21 & 0.55 & 337.4\\
             0.08 & 3.13 & 0.53 & 317.4\\
             0.35 & 1.00 & 2.29 & 180.0\\
  \end{tabular}
 \end{ruledtabular}
\label{tab:1}
\end{table}

\begin{figure}
 \epsfxsize=8.5cm
 \centerline{\epsffile{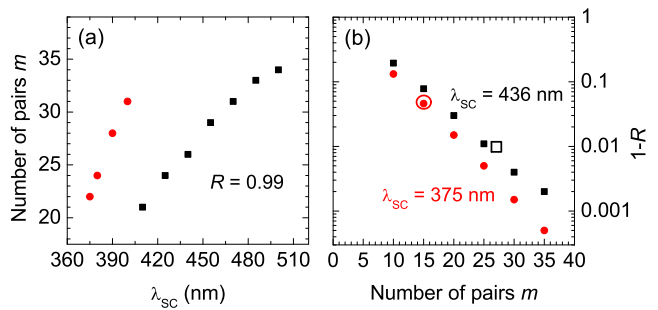}}
 \caption{\label{fig:2} (Color online)  Calculated properties of ZnO-based DBRs type 1 (squares) and 2 (circles). (a) Number of mirror pairs $m$ required to achieve  $R=0.99$ as a function of $\lambda_\text{SC}$. (b) Variation of reflectivity with mirror pair number $m$ at fixed $\lambda_\text{SC}$. Experimental results for DBRs in Fig.~\ref{fig:1} are marked by open symbols.}
\end{figure}

With the help of the RI dispersion curves, we are able to explore theoretically the
potential of DBRs type 1 and 2. Results of respective calculations are
summarized in Fig.~\ref{fig:2}. In Fig.~\ref{fig:2}(a), the number of
mirror pairs $m$ required for $R = 0.99$ is depicted as a function of
 $\lambda_\text{SC}$. At short
wavelengths, absorption of the high RI material limits the use of a
given DBR type. When aiming at longer wavelengths, an increasing
number of mirror pairs is necessary, which results from the
decreasing RI contrast between the layers. However, with less than
35 mirror pairs (equivalent to a DBR thickness $\lesssim 4$~$\mu$m), high
reflectivity ZnO-based DBRs can be realized which cover a wavelength
range from 375~nm up to 500~nm. For DBRs with a small
$\lambda_\text{SC}$, i.e., DBRs that do not work too far off the
excitonic resonance of the high RI material, the index contrast is
large enough to achieve very high reflectivities with only a
moderate number of mirror pairs. As shown in Fig.~\ref{fig:2}(b),
a 35 mirror pair type-1 DBR operating at
$\lambda_\textnormal{SC}=436$~nm can exhibit a reflectivity of
0.998, and a type-2 DBR with $\lambda_\text{SC}=375$~nm reaches
approximately the same value with only 30 mirror pairs. This makes the realization of monolithic
(Zn,Cd)O and ZnO-VCSELs feasible.

\begin{figure}
 \epsfxsize=7cm
 \centerline{\epsffile{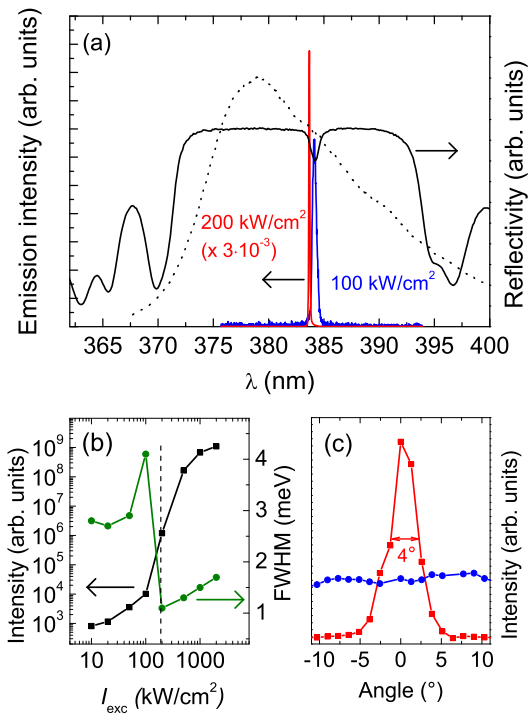}}
 \caption{\label{fig:3} FIG. 3: (Color online) Optical properties of a VCSEL device based on type-2 DBRs.
 (a) Normal incidence reflectivity (right axis), surface emission (left axis) for $I_\textnormal{exc}=100$~kW/cm$^2$ and $I_\textnormal{exc}=200$~kW/cm$^2$, and low-density MQW edge PL (dotted curve).
 (b) Integrated surface emission intensity (squares, left axis) and FWHM linewidth
 (circles, right axis) as a function of $I_\textnormal{exc}$. The lasing threshold $I_\textnormal{th}=190$~kW/cm$^2$ is marked by a dashed line. (c) Angle dependent,  spectrally integrated emission intensity at $I_\textnormal{exc}=100$~kW/cm$^2$ (circles) and $I_\textnormal{exc}=200$~kW/cm$^2$ (squares).}
\end{figure}

Based on the data and calculations of Figs.~\ref{fig:1} and
\ref{fig:2}, we have defined a design for a ZnO VCSEL. We use type-2 DBRs with 28 and
25 layer pairs as lower and upper mirrors,
respectively. The individual layer thicknesses are increased
slightly to shift $\lambda_\textnormal{SC}$ to $~$380~nm. A total
number of eight 4~nm wide ZnO/Zn$_{0.92}$Mg$_{0.08}$O QWs are
positioned near the electric field antinodes of a
Zn$_{0.92}$Mg$_{0.08}$O cavity with a total thickness of 2$\lambda$.
Correct MC growth is confirmed by normal incidence large-area ($1$~mm$^2$)
reflectivity measurement. As can be seen in Fig.~\ref{fig:3}(a) (black curve, right
axis) $\lambda_\textnormal{SC}$ is close to the desired position and a
reflectivity dip due to the cavity is observed at $384$~nm. The spectral
width of the cavity mode $\Delta E = 7$~meV corresponds to a quality factor
of $Q \approx 430$, though this is only a lower limit to $Q$, since
inhomogeneous broadening is known to play a role.

The device is optically excited by an
excimer laser pumped dye laser with a pulse duration of 20~ns
focused to a 60 $\mu$m diameter spot. At the excitation wavelength
$\lambda_{\textnormal{exc}}=364$~nm, reflectivity of the DBR is $20\,\%$ and absorption is small. Figure~\ref{fig:3}(a) (left
axis) shows surface cavity emission spectra at excitation densities of
$I_\textnormal{exc}=100$~kW/cm$^2$ and 200~kW/cm$^2$, as well as the low-density multiple QW (MQW) photoluminescence (PL) collected from the
sample edge (dotted curve) as reference. At low temperature (7~K, not shown), the full width at half maximum (FWHM) of the MQW PL is 10 meV implying that optical quality of the MQWs did not deteriorate notably by growth on the $2.6$~$\mu$m
thick lower DBR. At room temperature, on which we focus here, the FWHM is $135$~meV and the peak wavelength is at $379$~nm. Previous studies on the laser action of such MQW structures in an edge-emitting geometry have revealed that optical gain emerges on the low-energy side of the spontaneous-emission PL band and is related to localized states \cite{Cui06}. We have set the center wavelength of the DBRs to this spectral range, i.e., far below the MQW absorption peak which is even high-energy shifted with respect to the PL maximum by some 10 meV. Polariton effects related to strong exciton-photon coupling are thus unlikely to play a major role.

Comparing the cavity emission at $I_\textnormal{exc}=100$~kW/cm$^2$ and 200~kW/cm$^2$, two strong
indications for a transition to VCSEL action are worth to notice:
First, the emission intensity increases by more than two orders of magnitude, and
second, its FWHM decreases abruptly by a factor of 4. The evolution of both quantities with excitation density is depicted in
Fig.~\ref{fig:3}(b), signifying a well-defined threshold. The slight shift observed between the emission peaks below and above threshold might be due to changes in the RIs. Polarization fields affecting the low-density PL are screened off at the excitation levels close to threshold.
To confirm the realization of a VCSEL, we recorded emission-angle resolved spectra using a
microscope objective with a 20x magnification and a numerical
aperture of N.A.~$=0.5$ [see Fig.~\ref{fig:3}(c)]. As expected for a
VCSEL \cite{Bajoni07}, the emission above threshold is restricted to a narrow cone
about the surface normal with an angular FWHM of 4~$^\circ$ only.
Additionally, we find that the emission above threshold becomes strongly linearly polarized, probably due to the anisotropy of the sapphire substrate.

In conclusion, we have shown that monolithically grown DBRs based on
Zn$_{1-y}$Mg$_y$O/Zn$_{1-x}$Mg$_x$O heterostructures can be used as
high reflectivity mirrors for ZnO and (Zn,Cd)O active-layer
MCs. In particular, we demonstrated the realization of a VCSEL operating at $\lambda=384$~nm at room temperature. Current work is directed towards the strong-coupling regime, which may be reached by lowering the temperature or adapting the MC design.

This work was supported by Deutsche Forschungsgemeinschaft within SFB 555.

\end{document}